\documentclass[floatfix, noshowpacs, preprintnumbers, twocolumn, amsmath, amssymb, aps, prl, superscriptaddress]{revtex4}

\pdfoutput=1

\usepackage{graphicx, xcolor}
\usepackage[caption=false]{subfig}
\usepackage{soul}
\usepackage{slashed}
\usepackage{bm} \usepackage{graphicx} \usepackage{amsmath}
\usepackage{amsthm,stmaryrd}
\usepackage{amssymb} \usepackage{physics} \usepackage{txfonts}
\usepackage{color}
\usepackage{comment}
\usepackage{hyperref}

\usepackage{mathrsfs}
\usepackage{float}
\usepackage{array}

\usepackage{setspace}

\usepackage{pythonhighlight}

\begin{document}

\title{Teaching quantum information technologies and a practical module \\for online and offline undergraduate students}

\author{Hao Tang}
\altaffiliation{htang2015@sjtu.edu.cn}
\affiliation{Center for Integrated Quantum Information Technologies (IQIT), School of Physics and Astronomy and State Key Laboratory of Advanced Optical Communication Systems and Networks, Shanghai Jiao Tong University, Shanghai 200240, China}
\affiliation{CAS Center for Excellence and Synergetic Innovation Center in Quantum Information and Quantum Physics, University of Science and Technology of China, Hefei, Anhui 230026, China}
\altaffiliation{htang2015@sjtu.edu.cn}

\author{Tian-Yu Wang} 
\affiliation{Center for Integrated Quantum Information Technologies (IQIT), School of Physics and Astronomy and State Key Laboratory of Advanced Optical Communication Systems and Networks, Shanghai Jiao Tong University, Shanghai 200240, China}
\affiliation{CAS Center for Excellence and Synergetic Innovation Center in Quantum Information and Quantum Physics, University of Science and Technology of China, Hefei, Anhui 230026, China}

\author{Ruoxi Shi} 
\affiliation{Center for Integrated Quantum Information Technologies (IQIT), School of Physics and Astronomy and State Key Laboratory of Advanced Optical Communication Systems and Networks, Shanghai Jiao Tong University, Shanghai 200240, China}
\affiliation{CAS Center for Excellence and Synergetic Innovation Center in Quantum Information and Quantum Physics, University of Science and Technology of China, Hefei, Anhui 230026, China}

\author{Xian-Min Jin} 
\altaffiliation{xianmin.jin@sjtu.edu.cn} 
\affiliation{Center for Integrated Quantum Information Technologies (IQIT), School of Physics and Astronomy and State Key Laboratory of Advanced Optical Communication Systems and Networks, Shanghai Jiao Tong University, Shanghai 200240, China}
\affiliation{CAS Center for Excellence and Synergetic Innovation Center in Quantum Information and Quantum Physics, University of Science and Technology of China, Hefei, Anhui 230026, China}

\begin{abstract}
{\it Quantum Information Technologies and a Practical Module} is a new course we launch at Shanghai Jiao Tong University targeting at the undergraduate students who major in a variety of engineering disciplines. We develop a holistic curriculum for quantum computing covering the quantum hardware, quantum algorithms and applications. The quantum computing approaches include the universal digital quantum computing, analog quantum computing and the hybrid quantum-classical variational quantum computing that is tailored to the noisy intermediate-scale quantum (NISQ) technologies nowadays. Besides, we set a practical module to bring student closer to the real industry needs. The students would form a team of three to use any quantum approach to solve a problem in fields like optimization, finance, machine learning, chemistry and biology. Further, this course is selected into the Jiao Tong Global Virtual Classroom Initiative, so that it is open to global students in Association of Pacific Rim Universities at the same time with the offline students, in a specifically updated classroom. The efforts in curriculum development, practical module setting and blended learning make this course a good case study for education on quantum sciences and technologies.  
\end{abstract}

\maketitle

The research on quantum sciences and technologies have been emphasized at a national strategic level in many countries\cite{USquantum, Europe, China}. For instance, the USA signed the National Quantum Initiative Act\cite{USquantum} into law. It calls on different domestic government organizations to well cooperate on the research and development for a wide scope of quantum technologies. The European Union issued a `Quantum Manifesto' and announced its 1 billion-Euro Quantum Technologies Flagship\cite{Europe}, as they suggest the second quantum revolution is arriving and they aim to make Europe at a leading place in that era. 

The quantum sciences and technologies are generally regarded to include the three main fields, namely, quantum computing, quantum communication and quantum metrology. Micius Quantum Prize\cite{MiciusPrize}, an award named after the Chinese ancient philosopher and scientist ($\sim$400 B.C), has been launched to award the worldwide renowned quantum scientists. The Micius Prize also follows the above three categories, that is, being awarded to the field of quantum computation in 2018, quantum communication in 2019, quantum metrology in 2020, and to be announced on quantum computing again in 2021. Quantum computing utilizes the quantum mechanics to make more competitive computing frameworks as comparing to the classical computers, and hence it is hoped to be applied to wide fields that require powerful computing resources. Quantum communication that suggests more secure communication, as well as quantum metrology that suggests more accurate metrology and promotes fundamental physics research, also attracts researchers from multidisciplinary research background. 

In the upcoming decades, there is a huge demand for engineers and technicians who know how to use quantum sciences and technologies in their own fields. However, the current workforce is largely lagging behind that target, and hence the need for education on quantum sciences and technologies is severe. The US National Science Foundation (NSF) and the White House Office of Science and Technology Policy collaborate to develop quantum education for k-12 classrooms\cite{USk12}, which means teaching students from 5-6 years old to 17-18 years old. Even for undergraduate and graduate education, which is more urgent as it is more directly related to cultivating future engineers, the teaching for quantum sciences and technologies are not very common. Taking Shanghai Jiao Tong University as an example, this top-five university in China has not open courses on quantum computing in any department until very recent years. 

\begin{figure*}[ht!]
\includegraphics[width=0.85\textwidth]{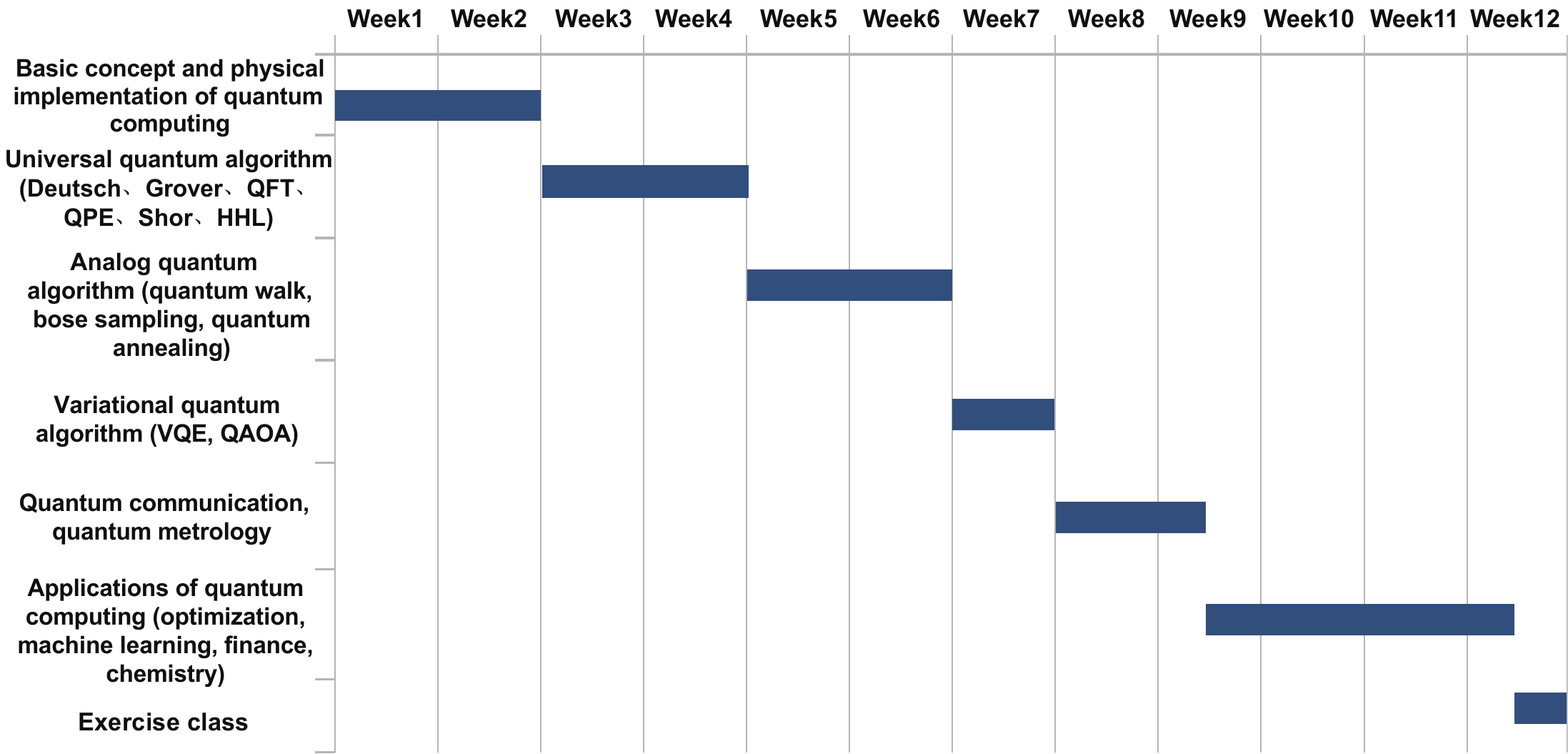}
\caption{\textbf{The Gantt chart for the teaching schedule.} The 12-week teaching curriculum includes the concepts, hardware, algorithms and applications of quantum computing, as well as a basic introduction on quantum communication and quantum metrology.  }
\label{fig:Concept}
\end{figure*}

In this letter, we share our experience in teaching {\it Quantum Information Technologies and a Practical Module}\cite{CourseLink}, a new course we launch at Shanghai Jiao Tong University targeting at the undergraduate students who major in a variety of engineering disciplines. This course includes a 3-credit teaching course and a 1-credit practical module. We develop a holistic curriculum for quantum computing covering the quantum computing hardware, quantum computing algorithms and applications, as well as a basic introduction on quantum communication and quantum metrology. The quantum computing approaches include the universal digital quantum computing, analog quantum computing and the hybrid quantum-classical variational quantum computing that is tailored to the noisy intermediate-scale quantum (NISQ) technologies nowadays. Besides, our 1-credit practical module brings student closer to the real industry needs. The students would form a team of three to use any quantum approach to solve a problem in fields like optimization, finance, machine learning, chemistry and biology. Further, this course is selected into the Jiao Tong Global Virtual Classroom initiative\cite{GVC}, so that it is open to global students in Association of Pacific Rim Universities\cite{APRU} at the same time with the offline students, in a specifically updated classroom. All these aspects require the teachers and teaching assistants to take relevant training. The efforts in curriculum development, practical module setting and blended learning make this course a good case study for education on quantum sciences and technologies.  

\begin{table*}[b!]
\caption{\textbf{A list of oral talk topics in class}}
\begin{center}
\begin{tabular}{p{8cm}<{\raggedright}p{0.9cm}<{\raggedright}p{3.5cm}<{\raggedright}p{0.9cm}<{\raggedright} }
\hline\noalign{\vskip 0.5mm}
		\hline\noalign{\vskip 0.14cm}
Topic & Class & Basic knowledge & Goal\\ [0.1cm]
\hline

How to make quantum gates using semiconductor qubits? & 3 & semiconductor qubits and CNOT gate & I \\ [0.6cm]

How to make quantum gates using ion trap qubits? & 3 & ion trap qubits and CNOT gate & I\\  [0.6cm]

We do use QFT and QAE for real applications. Can you please tell what the roles they play in these tasks? & 6 & Quantum Fourier transform  & II  \\  [0.6cm]

How do those alternatives to conventional QAE work and what are their advantages? & 6 & Quantum amplitude estimation & III \\  [0.6cm]

How to construct the quantum PCA? What is the function for each component of its circuit? Are there improved structures for quantum PCA? & 8 & HHL algorithm & III \\  [1.1cm]

What is quantum tensor network and how to theoretically and graphically study them? & 8 & Quantum Principal Component Analysis & III \\ [0.6cm]

Please explain what are the Hamiltonian matrices for various problems. & 9 & quantum walk and quantum simulation & II \\  [0.6cm]

What is quantum supremacy and what quantum algorithms can show that? & 10 & Boson sampling & II \\ [0.6cm]

What is Quantum Volume and how to measure it? & 10 & Boson sampling & II \\  [0.3cm]

What are photonic Ising machines, and why they can implement quantum annealing process? & 11 & Quantum annealing & I\\ [0.6cm]

What are the mechanisms of those cluster-state photonic computers? & 11 & quantum computing using photonics & I \\ [0.6cm]

The blockchain with something quantum & 12 & fast hitting & II\\ [0.3cm]

How to adapt VQE to calculate the higher energy states? & 13 & Variational Quantum Eigensolver & III\\ [0.6cm]

Quantum computing for chemistry: vibrational spectra using Gaussian Boson Sampling. & 13 & Gaussian Boson Sampling & II\\ [0.6cm]

How to use QAOA to do quantum factoring? & 14 & Quantum Approximate Optimization Algorithm & II\\ [0.6cm]

The continuous-time quantum walk is raised to implement a QAOA instead of the gate model. & 14 & Quantum walk and QAOA & III \\  [0.6cm]

QGAN and its application & 18 & Quantum machine learning & III\\ [0.6cm]

Quantum reinforcement learning and its application & 18 & Quantum machine learning & III\\ [0.6cm]
\hline\noalign{\vskip 0.5mm}
		\hline\noalign{\vskip 0.14cm}	
\end{tabular}

\begin{tabular}{p{13cm}<{\raggedright}}
Goal I: We emphasize hardware and physics for quantum devices. \\
Goal II: We hope students can digest news and analyze it with quantum knowledge.\\
Goal III: We encourage students to go deeper than textbooks and explore the frontier fields.\\
\end{tabular}

\end{center}
\end{table*} 

\section*{Curriculum development for a holistic view of the field}

Many students begin to enter the field of quantum computing by learning Nelson and Chuang's book entitled {\it Quantum Computation and Quantum Information}\cite{Chuang2010}. This is one of the most representative books on quantum computing, together with another book by Benenti, Casati and Strini entitled {\it Principles of Quantum Computation and Information}\cite{Benenti2004}. The latter is even more accessible as it is specifically written for undergraduate students in physics, mathematics and computer sciences, without too much requirement on prior knowledge on either quantum mechanics or classical compute sciences. We recommend both books to students in our class, especially the second one, as the students who have selected our course in the past two rounds are majored in physics, mathematics, computer science, software, electronics, information security, telecommunication, artificial intelligence, automation control, mechanical engineering, finance, and biology, covering a wide range of engineering disciplines. 

However, the above two books were firstly published in 2000 and 2004, respectively. This means they have almost missed the quickly development stage for quantum computing in the past two decades. The quantum computing hardware is much more advanced and scalable. The universal quantum algorithms that the two books emphasize have been developed. For instance, the HHL algorithm\cite{HHL} raised in 2009 becomes an important algorithm for solving linear equations. Furthermore, the adiabatic quantum computing and the hybrid quantum-classical variational quantum computing are now being extensively researched as an alternative to universal quantum computing, and find many quantum advantages in various appplications. 

Thanks to the convenience in the Internet, a quickly growing number of new resources for learning quantum computing can be found and accessed. Among them, some are shared by researchers on a certain field, and some are provided by commercial companies, like D-Wave and IBM. There then arises a issue for the resources. They are segmented and not that systematic. The tutorials provided by commercial companies might focus on the quantum computing approach used in their own companies. For instance, D-Wave\cite{DWaveOcean} teaches quantum annealing, while IBM Qiskit\cite{Qiskit} tutorial mainly covers circuit-based quantum computing. 

In developing our course, we suggest that we need to build a holistic view for quantum computing. We should not be biased on any quantum computing approach. Instead, we shall teach Grover algorithm, quantum annealing, Variational Quantum Eigensolver (VQE)\cite{VQE}, and Quantum Approximate Optimization Algorithm (QAOA)\cite{QAOA}, and show to students how to apply each of these different quantum algorithms on the same quadratic unconstrained binary optimization (QUBO) task. Through such training, students would have a better understanding of each quantum algorithm and a better vision on how to select a proper quantum algorithm when dealing with a specific computing task. 

The teaching for this course lasts 12 weeks, and each week has teaching time lasting 4$\times 45$ minutes (see teaching schedule in Fig.1). We start with a brief history of computers, and some basic knowledge on computer sciences and quantum mechanics. We then introduce the hardware for universal quantum computers using different physical systems, including superconducting qubits, cold atom or ion trap, linear photonics and different spin systems. After that, we begin to teach universal quantum algorithms including the early ones and a few new algorithms using quantum Fourier transform such as HHL. We highlight three sets of tools for studying quantum computing, that is, the underlying physics and hardware, the analytic tool such as matrix or Dirac notations, and the geometric tool such as the Bloch sphere and the angles between vectors used for understanding the Grover search. Especially, linear algebra with matrix calculation is the method we mainly use, without heavy quantum mechanics or classical computation teminologies.

\begin{figure}[b]
\includegraphics[width=0.48\textwidth]{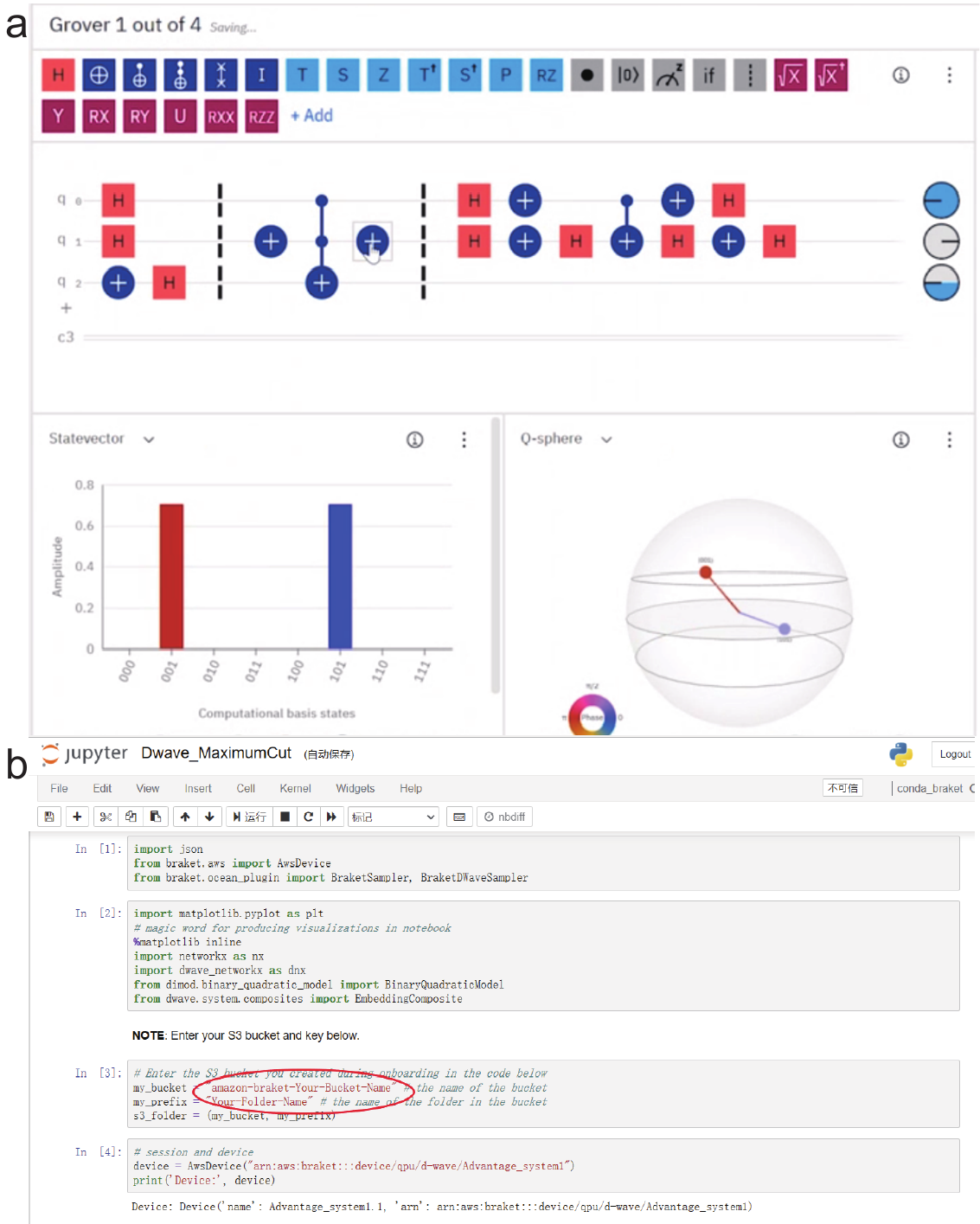}
\caption{\textbf{Interfaces for the quantum cloud platforms used in class.} (\textbf{a}) The graphic interface for universal quantum circuits at IBM Qiskit. (\textbf{b}) The interface for a Jupyter notebook that runs a Max-Cut problem on D-Wave quantum annealer.}
\label{fig:strutturaChip}
\end{figure} 

Apart from teaching universal quantum computing, we also teach students analog quantum computing and variational quantum computing. We teach continuous-time quantum walks, a useful tool for analog quantum computing, and show our experimental achievement using integrated photonic chips to make large-scale spatially two-dimensional quantum walks\cite{Tang2018}, as well as the application of quantum walks with quantum speedup advantages such as in quantum fast hitting algorithms\cite{Tang2018b}. The teaching of quantum walks have further meanings in at least three aspects. Firstly, we show that a combination of quantum walk and classical random walk to make a quantum stochastic walk\cite{Whitfield2010} would be useful for quantum simulation of the open quantum systems that widely exist in the reality. Secondly, we extend single-particle quantum walk into multiple-particle quantum walks, and show different behaviors of fermions, bosons and indistinguishable particles. With these knowledge it is much easier for students to understand Boson sampling\cite{BosonSampling}, a very popular task to show impressive quantum advantage, and a further useful variation into Gaussian Boson sampling\cite{GBS}. Thirdly, an understanding of the difference between quantum walk and classical random walk later help the students to learn how the quantum annealing differentiates from the classical simulated annealing. This is because quantum annealing\cite{QuantumAnnealing} or adiabatic quantum computing (AQC)\cite{AQC} can be viewed as an analog quantum evolution similar to quantum walk, but with a time-dependent Hamiltonian.

We teach variational quantum computing after quantum annealing and AQC. We follow that sequence not only because the variational quantum algorithms such as VQE and QAOA were raised many years after quantum annealing and AQC. Furthermore, QAOA, as an adapted adiabatic optimization algorithm that can be implemented in universal quantum circuits, is essentially inspired from AQC. Therefore, AQC sets as the prior knowledge for studying QAOA. On the other hand, variational quantum computing like VQE and QAOA inspires the concept and method of parameterized quantum circuit (PQC)\cite{PQC}, which is then widely used for quantum machine learning. Therefore, it is quite natural to proceed to the session on quantum optimization using different quantum algorithms, and the session on quantum machine learning with a quantum feature map using PQC. 

Throughout the course, we raise different oral talk topics for the students (see Table I). The students' grade for the 3-credit teaching course includes three major parts, the in-class performance and homework (30\%), the oral talk (35\%) and the final-term exam (35\%). The setting for oral talks is a very helpful way to extend the curriculum to a more comprehensive one, with the contribution from the students themselves according to their interest and strength. Some of the oral topics ask students go deeper into some quantum hardware, so that students from physics department always would like to go for them, while some other oral topics for quantum machine learning methods draw lots of interests from the computer science students. In another sense, the setting of oral talks also helps to relieve the discrepancy of students' prior knowledge and hence is very necessary for courses open to multidisciplinary students. It is also worth to mention that we do not list all oral talk topics at once, but announce two or three topics in every class with the topics related to what we learn in class. This helps students gain better understanding of the meaning of the oral talk topics.

\begin{table}[b!]
\caption{\textbf{Some practical projects and required quantum techniques}}
\begin{center}
\begin{tabular}{p{4.6cm}<{\raggedright}p{3.9cm}<{\raggedright}}
\hline\noalign{\vskip 0.5mm}
		\hline\noalign{\vskip 0.14cm}
Project topic & Quantum technique \\ [0.1cm]
\hline 

Protein folding  & Quantum annealing/VQE \\ [0.3cm]

Medical image classification & Quantum autoencoder\\ [0.3cm]

Computer Vision for 3D phase matching & Quantum annealing \\ [0.6cm]

Quantum frustration for spin liquid &VQE\\[0.6cm]

Reinforcement learning & Variational parameterized quantum circuit\\ [0.3cm]

Natural language processing for music & Variational parameterized quantum circuit\\ [0.1cm]
\hline\noalign{\vskip 0.5mm}
		\hline\noalign{\vskip 0.14cm}		
\end{tabular}
\end{center}
\end{table}

\section*{Practical module: tailored to industry needs} 
Throughout the 3-credit teaching course, we emphasize a practical skill. Each time after learning a quantum algorithm, the students are asked to set up the quantum algorithm on their own in a quantum computing platform. Initially, for simple universal quantum algorithms, we use the graphic interface at IBM Qiskit (see Fig.2a), which is of fun and brings an easy start for students who had little coding experience. Gradually, we will use a Jupyter note book to show how to set quantum algorithms with programming. The Jupyter notebook also works for teaching other quantum approaches, for instance, operating quantum annealing on D-Wave hardware (see Fig.2b). There is much work involved for our teaching assistants. They are expected to be those who know quantum computing well themselves, and are familiar with programming. The teaching assistant would prepare the Jupyter notebooks for quantum algorithms or case studies that will be used in class. They also address the questions that students raise on operating the programs, together with the teacher. As for the quantum computing cloud platform, we mainly use Amazon Braket as it has broad access to quantum annealers, universal quantum computers and classical computing resources. For the analog quantum computing part, we use the software FeynmanPAQS\cite{FeynmanPAQS} that we develop ourselves as a photonic analog quantum simulator. For the session on quantum machine learning, we use the framework DeepQuantum developed by a startup TuringQ. 

Our teaching content in the teaching course also shows a practical spirit. We have specific teaching sessions on quantum computing for optimization, machine learning, finance and chemistry. They are a good review and practise of all quantum algorithms that have been taught, and prepare the students with practical knowledge for the upcoming 1-credit practical module. In quantum optimization, we show how to map various real-life optimization problems such as Max-cut, vertex cover, knapsack, graph partitioning and maximal independent set into a QUBO Hamiltonian and then solve it using quantum algorithms. In quantum machine learning, we take quantum support vector machine (QSVM)\cite{QSVM} as an example, we show the original method of using a swap gate to create an inner product of the quantum state, and the derivation process is a good practise of deriving quantum states. We then show that the quantum feature map made from parameterized quantum circuits can also be used to form a QSVM\cite{QSVM2}. We explain main aspects that need to be considered for quantum machine learning, including data encoding, quantum feature maps and the parameter shift rule\cite{Crooks2019} as a preferred method for gradient descent in quantum circuits. For finance and chemistry, we show that these fields develop with computer sciences in history and now quantum approaches can help in various way. For instance, derivative pricing may need quantum amplitude estimation algorithm, portfolio optimization may resort to quantum annealing, and stock market prediction may involve some quantum machine learning. Similarly, molecular energy calculation, molecular docking and protein folding can also be solved using VQE, Gaussian Boson sampling or quantum annealing.   

Then we come to the 1-credit practical module during Week 13 and 16 after the 3-credit teaching course. The students are asked to form teams of three and solve a real problem using any quantum approach. This is quite similar to the mathematical modeling competition which many students have participated in before, so that they generally get used to this form very well. The topics are flexible and can be suggested by the students themselves. See Table II for a list of projects that the students have done for the practical module. 

There are some common requirements for all teams. Firstly, real input data is required. They can be real finance market portfolio, real protein structure, real medical image database (breast cancer, skin disease, et al), and so on. Students may try to reproduce the published work that has the real stock portfolio or protein data; They may also find data in Yahoo finance, protein bank, Github AI resources, et al. There is no strict lower bound for the data size. Secondly, at least one quantum approach shall be demonstrated. It will be even better if students compare quantum and classical approaches, or compare two or more quantum approaches (e.g. quantum annealing vs. QAOA, etc). They do not necessarily show quantum advantage over classical method, and of course it will be better and even publishable if they do show the advantage. Thirdly, the report should be well written in Latex and figures well presented.

\begin{figure*}[ht!]
\includegraphics[width=0.85\textwidth]{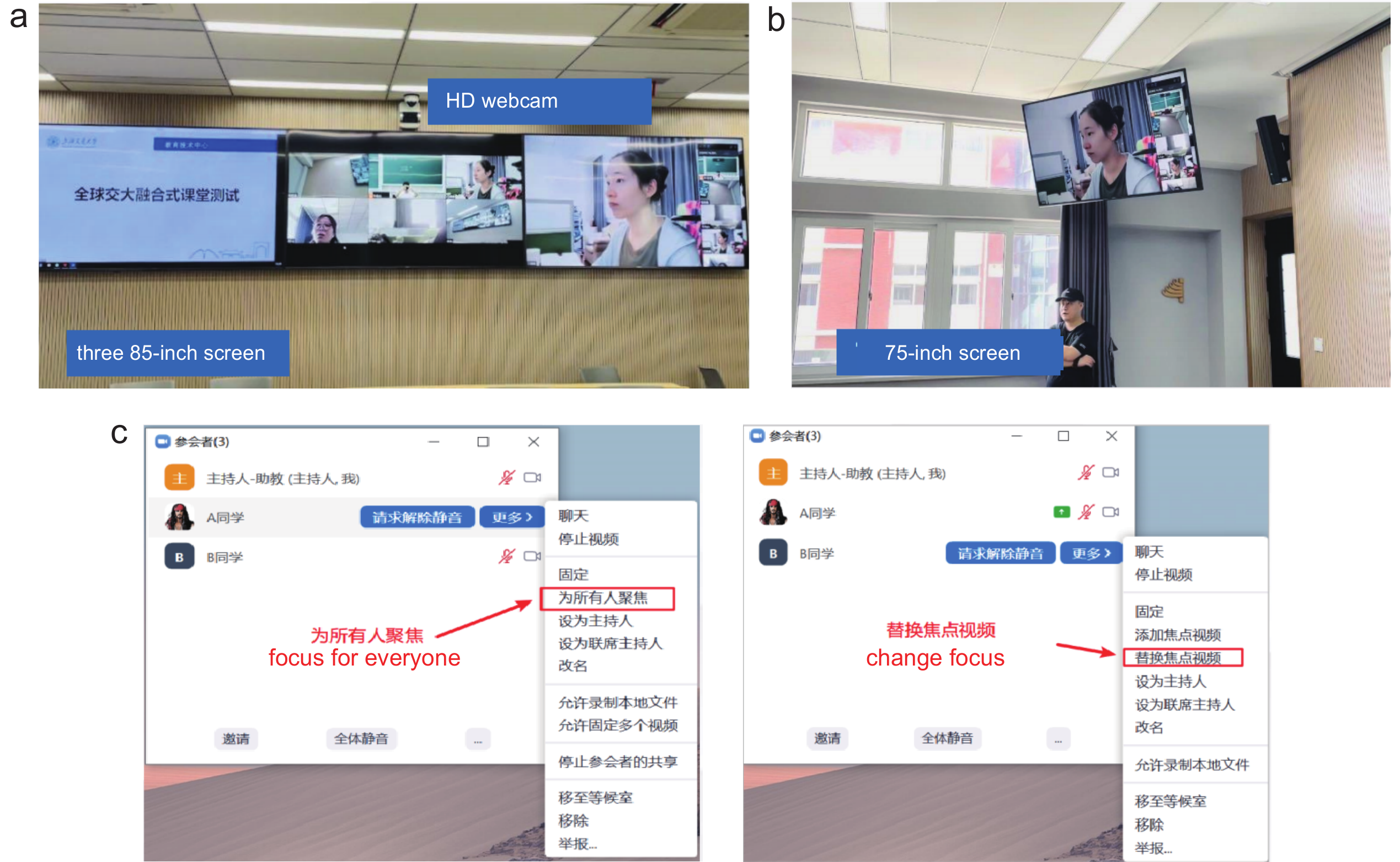}
\caption{\textbf{The setup for Jiao Tong Global Virtual Classroom.} (\textbf{a}) The three 85-inch screens on the back of the classroom. (\textbf{b}) The 75-inch screen in the front of the classroom. (\textbf{c}) The setting for the Zoom meeting by the teaching assistant.  }
\label{fig:Concept}
\end{figure*}

We give a careful thought on how to organize the practical classes during the four weeks. Students in one team will sit together for easier discussion. The class will include three main parts. Part One is the student report of their progress. For instance, in Week 13 they report on the topic and a basic plan for their project. In Week 14 one team member reports on how they find the data and encode the data in quantum approaches. In Week 15 another team member reports on how they set their quantum model, and in Week 16 the third team member reports on some initial results and how they plan to write their report. Part Two is some helpful expert tips given by researchers or engineers in the real fields or industries like quantum computing for ETF trading, drug design or computer vision. The knowledge goes deeper and more focused than what is taught in the teaching course. The Part Three are free Q\&A session, where the teacher, teaching assistants,and some guest expert speaker will help students sort out any questions they encounter in the project.

\section*{Jiao Tong Global Virtual Classroom} 

The Jiao Tong Global Virtual Classroom (GVC) Initiative\cite{GVC} encourages Shanghai Jiao Tong University (SJTU) faculty to establish collaborative online courses with partner universities across the globe, or share undergraduate/postgraduate-level courses to provide international students the opportunities to enhance cross-cultural competence, and deepen their understanding of the subject. 

Our course is selected into the Jiao Tong GVC Initiative. Since the autumn semester of 2021, it is open to global students in Association of Pacific Rim Universities at the same time with the offline students. Global students that have selected this course are undergraduate students in several top universities such as Hong Kong University, University College London and INSA Lyon, majored in several engineering subjects. 

This course now becomes an online-offline blended teaching. The mode of blended teaching was proposed a few decades ago and now an online and offline blended learning mode has become prevalent\cite{Sharma2010, Graham2006}. Especially, since the outbreak of coronavirus disease of 2019 (COVID-19), the demands for online learning soars and the online meeting software gets more and more convenient. For this course, we use Zoom meeting for online students to get into the class. We also use Zoom for weekly Q\&A session after class. 

In order to create a good virtual classroom for online students and ensure good interactions between online and offline students, the classroom has been updated specifically. The main difference is the additional use of four large screens and two HD webcams. As shown in Fig.3a and 3b, we have three 85-inch screens on the back of the classroom, and one 75-inch screen on the front-left side of the classroom. 

The three screens 85-inch screens are prepared for teachers. The left one will show the courseware controlled by teacher, which will also be shown at the front side of the classroom by a projector. The middle screen will show all the students. The right screen will show the same sight as the middle one but when teacher interacts with students online, this screen will show the close-up view of them (See Fig.3a). The 75-inch screen in the front is set for the offline students. It is set as the same with the right screen on the back, so that students in classroom can interact with those online face to face. 

While the screens help teachers and students in the classroom interact with the offline students, the two HD webcams are set for online students. One webcam captures the view of the front, including the teacher desk, projection of teaching slides and the blackboard. Especially, it can track the teacher so that when teacher writes on the blackboard it will give the writing a close-up view automatically and everyone can see the writing clearly. The other webcam shows a panoramic view of the class with every offline student shown in the scene. In addition, the audio equipment in the classroom is well set to ensure online students can hear the voice in class clearly. 

There is extra work and training for the teaching assistant (TA) than in usual classes. The TA needs to help set and control the Zoom account during the class. We need three Zoom accounts for the class. The first one is for teachers to show their courseware and the webcam view of the teacher. The second is connected to the HD webcam at the front of the classroom for the panoramic view of the class. These two accounts are set before class and don’t need to be managed during the class. The last one is controlled by the TA. When teacher asks student A to answer a question online or student A online put his/her hands up, TA will set A’s windows as ‘focus for everyone’ to give a close-up view of A for people in the classroom. If teacher wants student B to answer the next part of the question, TA will select ‘replace focus video’ to give a close-up view of B (See Fig.3c). If teacher wants both student A and B to have a discussion, TA will select ‘add focus video’ to give a close-up view of both A and B. After interaction, TA will reset all the settings. In the meeting, TA is the host and teacher would be co-host if needed. 

In our class for quantum information technologies, such settings for the classroom hardware and Zoom meeting ensures effective interaction between each pair of parties among the three, the teacher, the offline students and the online students. Meanwhile, all people, including teachers, TAs, online and offline students, have joined one discussion group on Wechat. During and after class, the students actively share their progress on quantum algorithms, and TA shares the link to some practical Jupyter notebooks in the Wechat discussion group. This works better than Canvas as the former provides an easier access on each's mobilephone. 

\section*{Discussion}

We manage to launch the course {\it Quantum Information Technologies and a Practical Module} owing to many aspects. First of all, the background of the quick development of quantum technologies and strong demand on relevant education gives the fundamental motivation. Besides that, the Zhiyuan Honors Program in the university encourages setting new innovative multidisciplinary course modules, which gives the opportunity to launch this course. Furthermore, the strong engineering research atmosphere in the university and the advice from Academic Affairs Office help to strengthen the style of the course with emphasis on practical skills and a practical module for engineering application. 

The development for an updated, holistic and systematic curriculum on quantum sciences and technologies is very necessary and urgent. We are writing up our teaching content on quantum computing into a tutorial book. It will show links between different quantum approaches, and sections on quantum computing for optimization, machine learning, finance, chemistry and biology. The book is with a hope to inspire readers from different fields to use quantum computing more widely, which is consistent with the purpose of the course.

Furthermore, the involvement of more technical support is helpful for courses on quantum sciences and technologies. We have shown the use of quantum cloud platforms and the setup for Jiao Tong Global Virtual Classroom that ensure a good effect for international, multidisciplinary and practical studies. In the future, when various quantum hardware get more mature and scalable, they may more commonly appear as a teaching tool in class, and that will give a further considerable boost for the teaching on quantum sciences and technologies.  

In all, we show our efforts in curriculum development, practical module setting and blended learning for our new course, and hope this can be a useful case study for state-of-the-art education on quantum sciences and technologies.  

\section*{Acknowledgements}
The authors thank Zhiyuan College at Shanghai Jiao Tong University for supporting the establishment of this course, the Academic Affairs Office at   Shanghai Jiao Tong University for helpful advices, and the International Affairs Division at Shanghai Jiao Tong University for launching the Jiao Tong Global Virtual Classroom Initiative. This course is supported by a funding on course development from Zhiyuan College (WF620160207/001/008) and a funding from Jiao Tong Global Virtual Classroom Initiative (WF610561701/003/006).

\section*{Author Contributions}
H.T. and X.M.J. are teachers for the course reported in this letter. T.Y.W. and R.S. are the teaching assistants for the course. T.Y.W. takes more responsibilities on monitoring the Zoom account and operations for the virtual classroom, while R.S. is more responsible for preparing for the practical Jupyter notebooks and address programming problems. H.T. writes this letter with input from all authors.

\bigskip

\end{document}